\documentclass[%
 reprint,%
 amssymb, amsmath,%
 aip,cha,%
]{revtex4-2}
\usepackage{graphicx,float,subfigure}
\usepackage{docs}%
\usepackage{bm}%
\usepackage[colorlinks=true,linkcolor=blue,citecolor=blue]{hyperref}%
\expandafter\ifx\csname package@font\endcsname\relax\else
\expandafter\expandafter
\expandafter\usepackage
\expandafter\expandafter
\expandafter{\csname package@font\endcsname}%
\fi
\usepackage{xcolor}

\usepackage{ulem}

\begin{document}

\title{Analytical Model for Gaussian Disorder Traps  in Organic Thin-Film Transistor}%

\author{Qiusong Chen}%

\altaffiliation[Also at ]{School of Physics and Electronic Science, Guizhou Education University, 115 Gaoxin Road, Guiyang 550018, China}%
\affiliation{Department of Materials Science, Fudan University, Shanghai 200433, China}

\author{Juan E. Sanchez}%
\affiliation{DEVSIM LLC, PO Box 50096, Austin, TX 78763, USA}

\author{Dong Lin}%
\affiliation{College of Information Engineering, Jimei University, Xiamen 361021, China}%

\author{Yanlian Lei}%
\affiliation{School of Physical Science and Technology, Southwest University, Chongqing 400715, China}%

\author{Guodong Zhu}%
\email{gdzhu@fudan.edu.cn}
\affiliation{Department of Materials Science, Fudan University, Shanghai 200433, China}

\begin{abstract}
Structural defects and chemical impurities exist in organic semiconductors acting as trap centers for the excited states. This work presents a novel analytical model to calculate the trapping and detrapping rates between two Gaussian density of states. Miller-Abrahams rate and Fermi–Dirac statistics are employed in this model. The introduction of effective filled and empty sites for correlated bands greatly simplifies the expression of recombination rate. A technology computer-aided design simulator-DEVSIM was used to simulate the donor-like traps in an organic semiconductor DPP-DTT based thin-film transistor, showing good agreement with the measured transfer characteristic. 
\end{abstract}

\date{\today}%
\maketitle


\section{Introduction}
Organic semiconductors are a class of promising materials for high-efficiency \cite{RN139}, low-cost \cite{RN1500},  flexible \cite{RN546}, and multi-functional \cite{Pandey} electronic devices.  
The trap states induced by the chemical impurities and structural defects in organic semiconductors play a critical role in electric performance. 

For example, chemical impurities have been doped in host materials as guest molecules to improve the electroluminescence in light-emitting diodes \cite{RN1938,RN1939} and to enhance the light absorption efficiency in dye sensitized solar cells \cite{RN1937,RN194}.
Other reports directly utilized the physics of trap states to build memories and sensors \cite{RN1934,RN1522,RN1861}. 

However, the applications of organic semiconducting materials in different areas are restricted by the trap states for the following two reasons: 1) the traps can serve as unintentional recombination centers in light-emitting diodes and solar cells \cite{RN301,RN308}, 2) carrier transport is also localized by chemical impurities and structural defects, resulting in carrier mobilities far below the theoretical prediction \cite{RN1906,RN1907}.
The study of trap related mechanisms is therefore crucial in extracting theoretical performance limits of organic semiconductors. Still many questions and challenges need to be addressed before engineering high-performance electronic devices.

In order to clarify the underlying mechanisms of trap related physics, several numerical methods have been proposed. 
The simplest approach included several discrete energy depths for traps, assuming that the trapping and detrapping rates are standard first order deferential equations with time\cite{RN1941,RN1894}. However, the trap states in organic semiconductors are dominated by continuous energy distributions \cite{RN1926,RN1861}.
Several works employed Gaussian distribution of exponential functions for trap levels to analyze the field-induced detrapping and thermal aging effects in organic semiconductors \cite{RN1865,RN1894,RN1920}. These works were mostly based on the master equation approach or Monte Carlo simulation\cite{RN1946,RN1865}. 
However, both approaches are time-consuming methods for device simulation \cite{RN1865}.
Hence, a more efficient and versatile approach is desired to explore the trap related kinetics.

In this study, we demonstrate a new method to analyze the charge generation (trapping) and recombination (detrapping) rate between two Gaussian density of states (DOS), by combining the Fermi–Dirac statistics with Miller-Abrahams equation \cite{RN1922}
Two variables of effective filled site (EFS) and effective empty site (EES) are defined in the expression of detrapping rate to simplify the algorithm complexity. Then, this method is applied to the understanding of the highest occupied molecular orbit (HOMO) and the donor-like trap (DLT) states in a Poly[2,5-(2-octyldodecyl)-3,6-diketopyrrolopyrrole-alt-5,5-(2,5-di(thien-2-yl)thieno [3,2-b]thiophene)] (DPP-DTT) \cite{RN449} based organic thin-film transistor(OTFT).
The simulated results demonstrated by a  technology computer-aided design (TCAD) platform of DEVSIM \cite{DEVSIM} show good agreement with the measured transfer characteristic.
This model can also be applied to other types of traps in organic electronic materials, such as acceptor-like traps.

\section{Modeling}
 
In both organic and inorganic materials, the energetic distribution of trap states is typically approximated by a Gaussian function with a standard energetic deviation or an exponential function with a modified characteristic temperature \cite{RN1861,RN1796,RN1920,RN1526,RN1926}. 
The Gaussian function is a better choice to describe the limited nature of trap density. 

Because of the less crystallized structure in organic materials, the energetic disorder of conductive bands in organic semiconductors are typically also given by Gaussian DOS \cite{Paasch}. Unlike the inorganic crystalline structure, there is no clear boundary between the allowed band and the forbidden band \cite{RN1925}. Boltzmann statistics is therefore not a proper approximation for the states in amorphous organic semiconductors \cite{RN1925} and Fermi–Dirac statistics should be employed to calculate the average number in a single-particle state \cite{RN1877}.

The charge transition model, including trapping and detrapping, between various energy levels in organic semiconductors must be treated differently from inorganic crystals. In the following context, we employ Gaussian DOS to describe energy statistics for both of HOMO and DLT , and then use Miller-Arbrhams rate and Fermi–Dirac statistics to describe the charge transition between these two energy levels. 

\subsection{The Gaussian DOS for HOMO and DLT}
\begin{figure}[H]
	\centering 
	\includegraphics[width=0.5\textwidth]{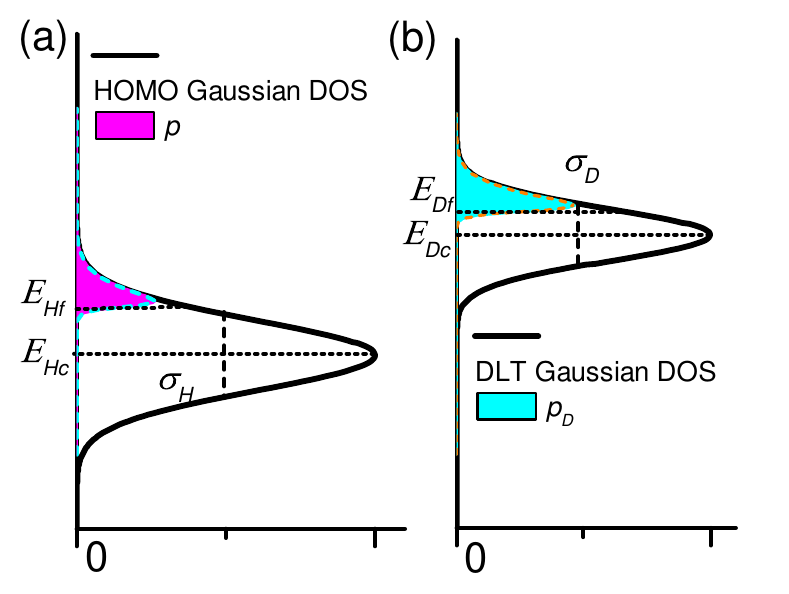} 
	\caption{The diagrams of filling states in Gaussian DOS for HOMO and DLT} 
	\label{Fig.DOS} 
\end{figure}
The Gaussian DOS for HOMO ($g_H$) of organic materials is as follows :
\begin{equation}
g_H(E_H,E_{Hc},\sigma_H)=\frac{N_H}{\sigma_H\sqrt{2\pi}}\exp(-\frac{(E_H-E_{Hc})^{2}}{2\sigma_H^{2}}) \label{HOMO DOS}
\end{equation}
where ${E_H}$ is an energy level of a specific state in HOMO. $N_H$, $\sigma_H$, and $E_{Hc}$ are the total density, the distribution width, and energy center of HOMO DOS, respectively. In the following content, we will use $g_H(E_H)$ as a shorthand for $g_H(E_H,E_{Hc},\sigma_H)$.

The black solid curve in Fig. \ref{Fig.DOS}(a) represents the Gaussian distribution for HOMO DOS and the empty sites in it.
In addition, a hole is defined to be the remaining delocalized positive charge after an electron escapes form a HOMO site, whose density is: 

\begin{equation}
	p=\int_{-\infty}^{\infty} g_H(E_H,E_{Hc},\sigma_H)(1-f(E_H,E_f)) dE_H \label{HOMO Holes}
\end{equation}
in which $f(E,E_f)$ is Fermi–Dirac statistics and it reads:
\begin{equation}
f(E,E_f)=\frac{1}{e^{\frac{(E-E_{f})}{k_{B}T}}+1} \label{Fermi}
\end{equation}
where $E_f$, $k_B$, $T$ are Fermi energy level, Boltzmann's constant, and the absolute temperature, respectively. 

It should be noted that, only in the equilibrium state, different bands in the system share the same Fermi level. Otherwise, in the non-equilibrium state, every band has its own Fermi level to describe the overall occupation of the sites \cite{RN1898}. In such condition, this energy level is termed as quasi Fermi level. As the trapping and detrapping processes typically occur in a non-equilibrium state, we should use quasi Fermi level of different bands in the following context, i.e. $E_{Hf}$ for HOMO and $E_{Df}$ for DLT.

The trapped charges consist of both donor-like and acceptor like states across the forbidden energy gap \cite{RN1773}. 
Then, the DOS for deep-level DLT is in the same manner of Gaussian distribution:
\begin{equation}
g_{D}(E_D,E_{Dc},\sigma_{D})=\frac{N_{D}}{\sigma_{D}\sqrt{2\pi}}\exp(-\frac{(E_D-E_{Dc})^{2}}{2\sigma_{D}^{2}}) \label{DLT DOS}
\end{equation}
where ${E_D}$ is an energy level of a specific state in DLT. $N_D$, $\sigma_D$, and $E_{Dc}$ are the total density, the distribution width, and energy center of DLT DOS, respectively. In the following content, we will use $g_{D}(E_D)$ as a shorthand for $g_{D}(E_D,E_{Dc},\sigma_{D})$.

The black solid curve in Fig. \ref{Fig.DOS}(b) represents the Gaussian distribution for DLT and the empty sites in it.
Since the DLT site is neutral if occupied by an electron and positively charged if empty, the density of trapped charge corresponds to the empty sites. Then the trap density reads:
\begin{equation}
	p_D=\int_{-\infty}^{\infty} g_{D}(E_D,E_{Dc},\sigma_{D})(1-f(E_D,E_{Df})) dE_D \label{DLT Holes}
\end{equation}

\subsection{Miller-Abrahams Equation}
To describe the transition rate of electrons from one site with energy level of $E_i$ to another site with $E_j$ , we employ Miller-Abrahams rate \cite{RN1922,RN1865}, which reads:
\begin{equation}
v_{ij}=v_0\exp(-2\frac{R_{ij}}{a_i}-\frac{E_j-E_i+|E_j-E_i|}{2k_BT}) \label{Miller-Abrahams}
\end{equation}
where $v_0$ is attempt frequency, $a_i$ is the localization scale of initial state, $R_{ij}$ is the distance between both states. 

\subsection{Trapping Process}
\begin{figure}[H] 
	\centering 
	\includegraphics[width=0.5\textwidth]{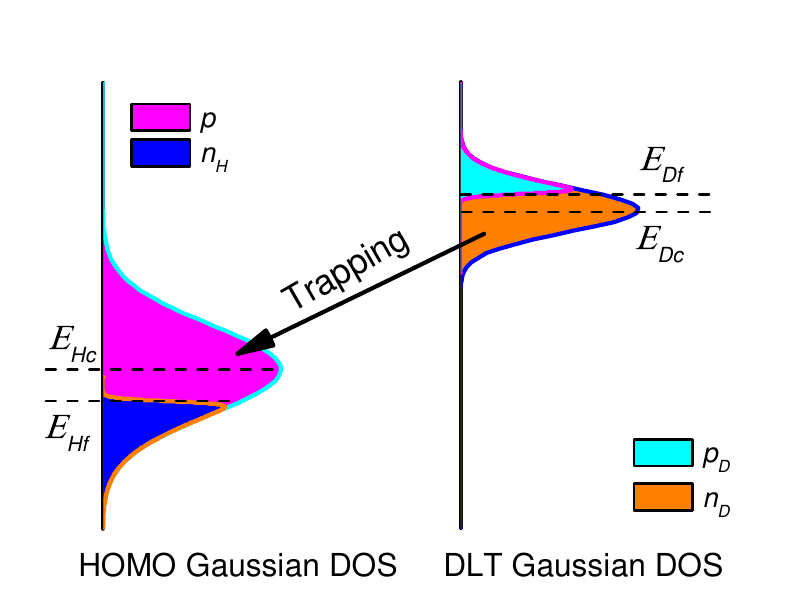} 
	\caption{The diagrams of electron's trapping process from the neutral sites in DLT to the empty sites in HOMO. $p$ and $n_H$ are the empty sites(holes) and the filled sites in HOMO DOS. $n_D$ and $p_D$ denote the filled and the empty sites in DLT DOS} 
	\label{Fig.Trapping} 
\end{figure}

 For the trapping process, the electron jumps from a neutral DLT site to an empty HOMO site. Here, we assume that all states of DLT are higher than HOMO, because Gaussian distributed DLT are deep-level bands \cite{RN1773}. 
 Then by integrating the transition rates of all electrons in DLT to all empty site in HOMO, we can get the total rate of trapping. Fig. \ref{Fig.Trapping} exhibits this process. Hence, the overall trapping rate reads:
\begin{equation}
\begin{aligned} 
	k_T=\iint_{-\infty}^{\infty}& v_{DH} g_{D}(E_D)f(E_D,E_{Df}) \\
	&g_H(E_H)(1-f(E_H,E_{Hf})) dE_D dE_H
\end{aligned} \label{Trapping Integral}
\end{equation}
where $E_D,E_H$ are two energy sites located on DLT and HOMO, respectively. $v_{DH}$ is Miller-Abrahams rate for electron transition form site of $E_D$ to $E_H$. It reads :
\begin{equation}
	v_{DH}=v_{0}\exp(-2\alpha_D R_{DH}) \label{Trapping Coefficient}
\end{equation}
where $\alpha_D$ is the inverse localization scale of DLT site, $R_{DH}$ is the average distance from HOMO sites to DLT sites.

After separating variables, the above expression can be simplified as:
\begin{equation}
k_T=v_{DH} n_Dp \label{Trapping Rate}
\end{equation}
in which $n_D$ is the filled sites in DLT.  It is illustrated as the orange area in Fig. \ref{Fig.Trapping} and reads:
\begin{equation}
\begin{aligned}
n_D &=\int_{-\infty}^{\infty} g_{D}(E_D,E_{Dc},\sigma_{D})f(E_D,E_{Df}) dE_D \\
    &= N_D - p_D \label{FS DLT}
\end{aligned}
\end{equation}

\subsection{Detrapping Process}
\begin{figure}[H]
	\centering 
	\includegraphics[width=0.5\textwidth]{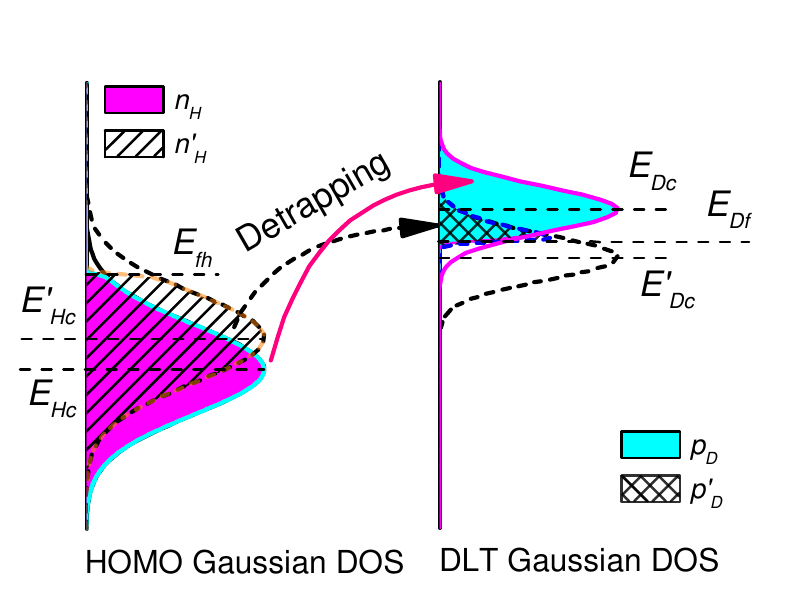} 
	\caption{The diagrams of electron's detrapping process from HOMO to the empty sites in DLT. $n_H$ is the actual filled sites and $n_H^{'}$ is the effective filled sites in HOMO.  $p_D$ is actual empty sites and $p_D^{'}$ is the effective empty sites in DLT} 
	\label{Fig.Detrapping} 
\end{figure}

For the detrapping process, the electron jumps from a filled HOMO site to an empty DLT site as indicated by the red solid arrow in Fig. \ref{Fig.Detrapping}. Then by integrating all electrons in HOMO to all empty sites in DLT, we can get the total rate of detrapping. So the detrapping rate reads:
\begin{equation}
\begin{aligned} 
k_D=\iint_{-\infty}^{\infty}& v_{HD} g_H(E_H)f(E_H,E_{Hf}) \\
&g_{D}(E_D)(1-f(E_D,E_{Df})) dE_H dE_D
\end{aligned} \label{Detrapping Integral}
\end{equation}
where $v_{HD}$ is Miller-Abrahams rate for electron's transition form the site of $E_H$ in HOMO to $E_D$ in DLT. $v_{HD}$ reads:
\begin{equation}
v_{HD}=v_0\exp(-2\frac{R_{DH}}{a_H} -\frac{E_D-E_H}{k_BT})\label{Detrapping Coefficient}
\end{equation}
where $a_H$ is the average localization scale of HOMO site.

Using the following replacement:
\begin{equation}
	v_{HD0}=v_0\exp(-2\alpha_H R_{DH})
\end{equation} 
the expression of Eq. \eqref{Detrapping Integral} can be rewritten as :
\begin{equation}
\begin{aligned} 
k_D= \iint_{-\infty}^{\infty} & v_{HD0}\exp(-\frac{E_D-E_H}{k_BT})g_H(E_H)f(E_H,E_{Hf})\\ 
& g_{D}(E_D)(1-f(E_D,E_{Df})) dE_H dE_D
\end{aligned} \label{Detrapping Integral2}
\end{equation}

\subsection{Effective Empty and Filled Sites}
Eq. \eqref{Detrapping Integral2} has one more exponential terms than Eq. \eqref{Trapping Integral}, but it can still be simplified by the method of separation of variables. Here, we define two parameters of $E^{'}_{Hc}$ and $E^{'}_{Dc}$ as effective energy centers for HOMO and DLT DOS, respectively. The expressions of two effective energy centers are:
\begin{equation}
	E^{'}_{Hc}=E_{Hc}+\frac{\sigma_H^{2}}{k_BT} \label{HEEC}
\end{equation}
\begin{equation}
E^{'}_{Dc}=E_{Dc}-\frac{\sigma_D^2}{k_BT}\label{DEEC}
\end{equation}
Fig. \ref{Fig.Detrapping} illustrate the energy shift of $E^{'}_{Hc}$ and $E^{'}_{Dc}$  with respect to $E_{Hc}$ and $E_{Dc}$, respectively.
By introducing effective energy centers for both HOMO and DLT levels, Eq. \eqref{Detrapping Integral2} can be rewritten as :
\begin{equation}
\begin{aligned} 
k_D=&c_{E} v_{DH0} \iint_{-\infty}^{\infty} g_H(E_H,E^{'}_{Hc},\sigma_H)f(E_H,E_{Hf})\\ 
& g_{D} (E_D,E^{'}_{Dc},\sigma_{D})(1-f(E_D,E_{Df})) dE_H dE_D
\end{aligned} \label{Datrapping Integral3}
\end{equation}
in which $c_{E}$ reads:
\begin{equation}
	c_E=\exp(\frac{\sigma_H^2+\sigma_{D}^2}{2k_B^2T^2}+\frac{E_{Hc}-E_{Dc}}{k_BT})
\end{equation}
Compared with Eq. \eqref{Trapping Integral}, Eq. \eqref{Datrapping Integral3} can also be simplified to a product form like Eqs. \eqref{Trapping Rate}:
\begin{equation}
k_D=c_{E} v_{DH0} n_H^{'}p^{'}_D \label{Detrapping Rate}
\end{equation}
in which $n_H^{'}$ reads:
\begin{equation}
n_H^{'}=\int_{-\infty}^{\infty} g_H(E_H,E^{'}_{Hc},\sigma_H)f(E_H,E_{Hf}) dE_H \label{HOMO EEHoles}
\end{equation}
and $p^{'}_D$ reads:
\begin{equation}
p_D^{'}=\int_{-\infty}^{\infty} g_{D}(E_D,E^{'}_{Dc},\sigma_{D})(1-f(E_D,E_{Df})) dE_D \label{DLT EEHoles}
\end{equation}

Because Eqs. \eqref{HOMO EEHoles} and \eqref{DLT EEHoles} are similar to Eqs. \eqref{FS DLT} and \eqref{DLT Holes}, respectively, we define $n_H^{'}$ as EFS for HOMO and $p^{'}_D$ as EES for DLT. 

So far, we used both Fermi-Dirac statistics and Miller-Abrahams equation to derive the trapping and detrapping rates of DLT in the Gaussian energetic disorder semiconductors. The calculation complexity caused by the barrier in the detrapping process was reduced by virtue of introducing EFS and EES for the correlated bands. Then the detrapping rate is simplified from a sophisticated integral expression (Eq. \eqref{Detrapping Integral}) to a concise product form (Eq. \eqref{Detrapping Rate}). The black dot arrow in Fig. \ref{Fig.Detrapping} indicates the simplified detrapping picture. In the following section, we will analyze the relationship between different parameters and variables in this model. Then, this model was employed to simulate a DPP-DTT  based transistor with a bottom-gate/ top-contact configuration. 

\section{Data ANALYSIS}
\subsection{Parameters and Variables}

\begin{figure}[H]
	\centering 
	\includegraphics[width=0.5\textwidth]{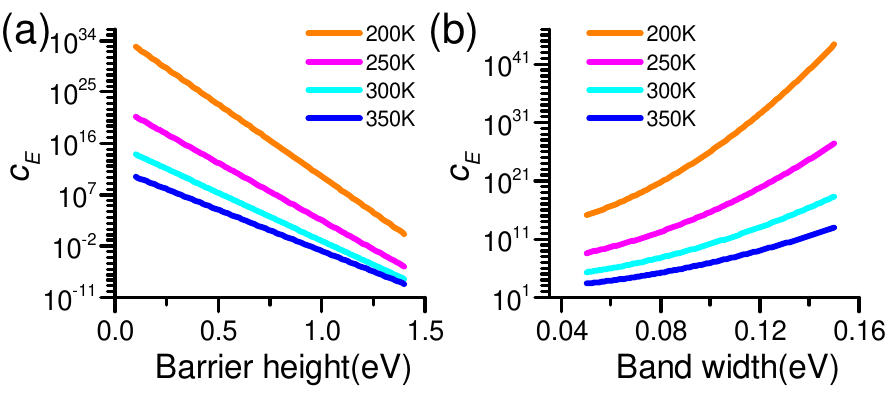} 
	\caption{(a) Dependencies of $c_E$ on barrier height ($E_{Dc}-E_{Hc}$) at various temperature. All Gaussian widths for these lines are fixed at 0.11 eV. (b) Dependencies of $c_E$ on one band width of Gaussian HOMO or DLT DOS at various temperature. The barrier height for these curves are set to 0.25 eV.} 
	\label{Fig.CoefficientC} 
\end{figure}
In the above content, the trapping and detrapping rates are reduced to the product forms in Eq. \eqref{Trapping Rate} and \eqref{Detrapping Rate}. The coefficient $v_{DH}$ for trapping in Eq. \eqref{Trapping Rate} reflects the impacts of orbit localization in DLT and spacial distance from HOMO to DLT. The other two variables are the filled electrons in DLT ($n_D$) and the empty sites in HOMO ($p$). But for the detrapping rate, there's an extra coefficient of $c_{E}$  in Eq. \eqref{Detrapping Rate}. Fig. \ref{Fig.CoefficientC} (a) and (b) present the dependence of $c_E$ on the barrier height and the band width of Gaussian DOS. 
In Fig. \ref{Fig.CoefficientC} (a), $c_E$ shows a dramatically exponential decay with respect to the barrier height between the Gaussian centers of HOMO and DLT. This feature intuitively describes the physical role of energy barrier.
However, there are two results to note: \textcircled{1}, most of $c_E$ values within the range shown in Fig. \ref{Fig.CoefficientC} is greater than 1; 
\textcircled{2} the range of $c_E$ decreases with device temperature for in both Fig. \ref{Fig.CoefficientC} (a) and (b). 
Both results are inconsistent with Miller-Abrahams rate in Eq. \eqref {Miller-Abrahams}, because the detrapping rate should be reduced by the energy barrier ($c_E < 1$) and enhanced with higher temperature.  
So, the attenuation effect of the potential barrier on detrapping process will act more on EES and EFS. in Eq. \eqref{Detrapping Rate}.

\begin{figure}[t]
	\centering 
	\includegraphics[width=0.5\textwidth]{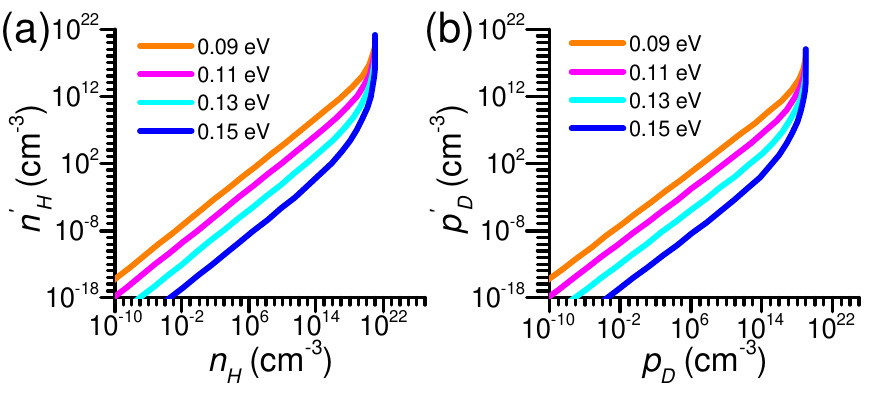} 
	\caption{(a) Dependencies of $n^{'}_H$ on $n_H$ of various band width for Gaussian HOMO DOS at $300 K$. (b) Dependencies of $p_D^{'}$ on $p_D$ for various band width of Gaussian DLT DOS at $300 K$. The total density for HOMO and DLT DOS are $1.2 \times 10^{21} cm^{-3}$ and $1.0 \times 10^{19} cm^{-3}$, respectively. }  
	\label{Fig.EffectiveSites} 
\end{figure}

Fig. \ref{Fig.EffectiveSites} presents the dependence of EES and EFS on the actual charge density of $n_H$ in HOMO and $p_D$ in DLT, respectively. These data are obtained by Paasch's method about the Gauss–Fermi integral \cite{Paasch}.
It can be observed from Fig. \ref{Fig.EffectiveSites} (a) that the $n^{'}_H$ relationship with the $n_H$ asymptotically reaches a slope equal to 1 in the log-log plot for small densities. And, the $p^{'}_D$ relationship with the $p_D$ presents the same manner. 
In this region, the values of $n^{'}_H$ and $p^{'}_D$ are $5 \sim 15$ orders of magnitude smaller than $n_H$ and $p_D$ , respectively. The magnitude of difference increases with the Gaussian DOS widths, because the energy shift of $E^{'}_{Hc}$ and $E^{'}_{Dc}$ increases with these two widths in Eqs. \eqref{HEEC} and \eqref{DEEC}. 
However, these magnitude differences decrease when $n_H$ and $p_D$ become saturated. So the slop of these curves becomes very steep in these regions in both Fig. \ref{Fig.EffectiveSites} (a) and (b). Such phenomenon reflects the saturation characteristic of the Gaussian distribution. Take the Gaussian HOMO DOS as example, the quasi Fermi level is the only variable of EFS' expression in Eq. \eqref{DLT EEHoles}. So, while the actual filled sites in HOMO DOS are saturated, the raised effective energy center of $E^{'}_{Hc}$ makes EFS increase obviously.

\subsection{TCAD Simulation for OTFT}
In organic semiconductors, the surface trap density is more prominent than the bulk trap density.
A device with active region on the surface is therefore an appropriate application to validate this method. And OTFT meats this criteria, because the conductive channel is located in the interface area of semiconductors.

In the following content, we demonstrate our model in a TCAD simulation of an OTFT device with the bottom-gate/top-contact configuration. 
The simulation parameters and device dimensions are listed in Table \ref{tab:table1}. The DEVSIM \cite{DEVSIM} TCAD simulator was employed to realize a finite volume method (FVM) analysis. The typical TCAD approach of drift-diffusion model solves three equations:

\paragraph{The Poisson equation:}
\begin{equation}
-\nabla (\varepsilon_r \varepsilon_0 \nabla \psi) =e \cdot (p-n+p_D) \label{Poisson}
\end{equation}
where $\psi$ is electrical potential, $\varepsilon_0$ is the permittivity of vacuum, $\varepsilon_r$ is the relative permittivity of the organic material, and $e$ is the elementary charge. 

\paragraph{Hole Continuity Equation:}
\begin{equation}
\frac{\partial p}{\partial t} =\nabla (p \mu_p \nabla\psi + D_p \nabla p)  + G_p \label{HoleContinue}
\end{equation}
where $\mu_p$, $D_p$ and $G_p$ are the mobility, diffusion coefficient and net generation rate for holes, respectively.

\paragraph{Electron Continuity Equation:}
\begin{equation}
\frac{\partial n}{\partial t} =\nabla (-n \mu_n \nabla\psi + D_n\nabla n  ) + G_n \label{ElectronContinue}
\end{equation}
where $\mu_n$, $D_n$ and $G_p$ are the mobility, diffusion coefficient and net generation rate of electrons, respectively.

To calculate the variance of DLT density over time, an extra equation for the net generation rate of DLT is solved simultaneously with the above three equations. The net generation rate of DLT reads:
\begin{equation}
\frac{\partial q_D}{\partial t} = k_T - k_D
\end{equation}
\begin{table}[H]
	\caption{\label{tab:table1} The parameters used to simulate the DPP-DTT based OTFT if not explicitly stated elsewhere. }
	\begin{ruledtabular}
		\begin{tabular}{lcrl}
			Model & Symbol &Value & Unit\\
			\hline
			Device temperature &	$T$ & $300$& K\\
			Total density for HOMO \cite{RN1873}  &	$N_H$ & $1.2\times10^{21}$ & cm$^{-3}$\\
			HOMO DOS width \cite{RN1877,RN1898} & $\sigma_H$ & 0.13 &eV\\
			HOMO DOS center \cite{RN1550} & $E_{Hc}$ & -5.2 & eV\\
			LUMO DOS center \cite{RN1550}&          & -3.5 & ev\\
			Total density for DLT \cite{RN1861} &	$N_{Ds}$ & $1.0\times10^{13} $& cm$^{-2}$\\
			DLT DOS width & $\sigma_D$ & 0.11 &eV\\
			DLT DOS center & $E_{Dc}$ & 3.81 &eV \\ 
			Trap thickness & $d_D$ & 3 & nm \\
			Average distance between \\HOMO and DLT & $R_{DH}$ & 5 & nm \\
			Localization length  \\for HOMO and DLT\cite{RN1865} & $a_i$ &  0.5 & nm\\
			Attempt frequency\cite{RN1865} & $v_0$ & 1  & s$^{-1}$\\
			DPP-DTT thickness &  & 40 & nm \\
			Insulator's thickness &  & 300 & nm \\
			OTFT channel length &  & 100 & $\mu$m \\
			Relative permittivity \\for DPP-DTT  & $\varepsilon_{r\_Semi}$ & 4 &  \\
			Relative permittivity \\for Insulator SiO$_2$  &   $\varepsilon_{r\_SiO_2}$ & 3.9 &  \\
			Thermionic emission velocity\\
			at source and drain \cite{RN1932} &  & $10^{7}$  & cm/s  \\
		\end{tabular}
	\end{ruledtabular}
\end{table}

As our method is based on the bulk density of DLT, we convert the total surface trap density to the total bulk density through an exponential decay relationship with depth:
\begin{equation}
N_D = \frac{N_{Ds}}{d_D} \exp (-\frac{x}{d_D}) \label{DensityTran}
\end{equation}
where $x$ is the depth from the insulator-semiconductor interface to the semiconducting layer, $d_D$ is the characteristic thickness of DLT. The values for these parameters are listed in Table \ref{tab:table1}. Integrating Eq. \eqref{DensityTran} on the depth in semiconductor layer, the sum of bulk density is consistent with surface density.

Here, we assume there is only one kind of trap. So, in Eq. \eqref{Poisson}, the total charges in Poisson's equation include three parts: holes, electrons and DLT. As the organic OTFT works in the accumulation regime for on-state and depletion regime for off-state, we take only the electron's transition between HOMO and DLT in consideration. Then, we can get the following two results: 
\begin{equation}
	G_p=-(k_T-k_D)
\end{equation}
\begin{equation}
	G_n=0
\end{equation}

For the carriers transport, we use the macroscopic
conductivity model for holes \cite{RN537, RN1907} and Poole–Frenkel model for electrons \cite{RN1773}. Then, a typical model for Schottky contacts is considered on source and drain \cite{RN1933}. The thermionic emission velocity for holes and electrons is listed in Table. \ref{tab:table1}. 
The initial equilibrium was solved while $k_T$ equal to $k_D$.


\section{Experiment and Simulation Results}
\begin{figure}[H]
	\centering 
	\subfigure{\includegraphics[width=0.5\textwidth]{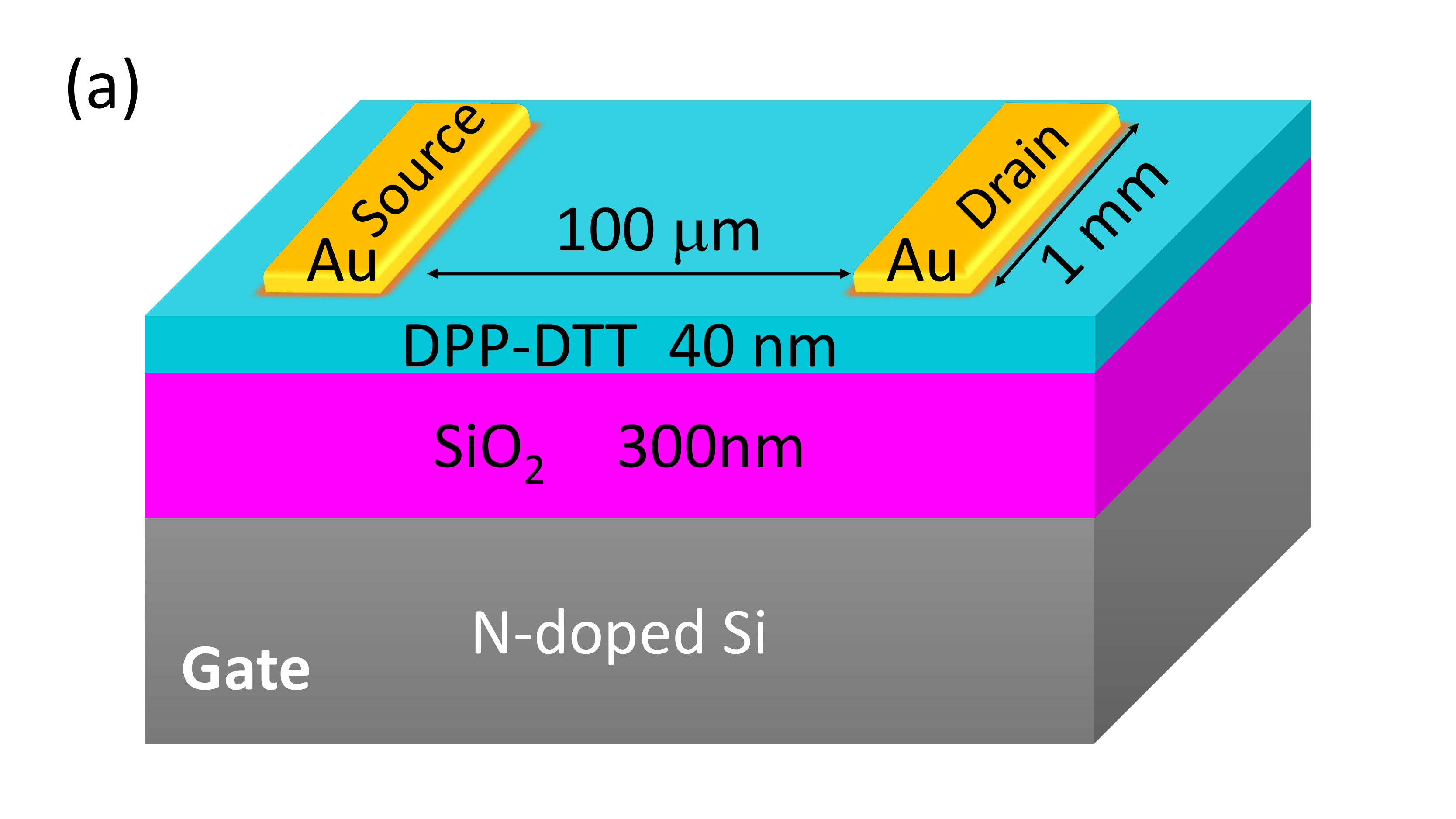}}
	\subfigure{\includegraphics[width=0.5\textwidth]{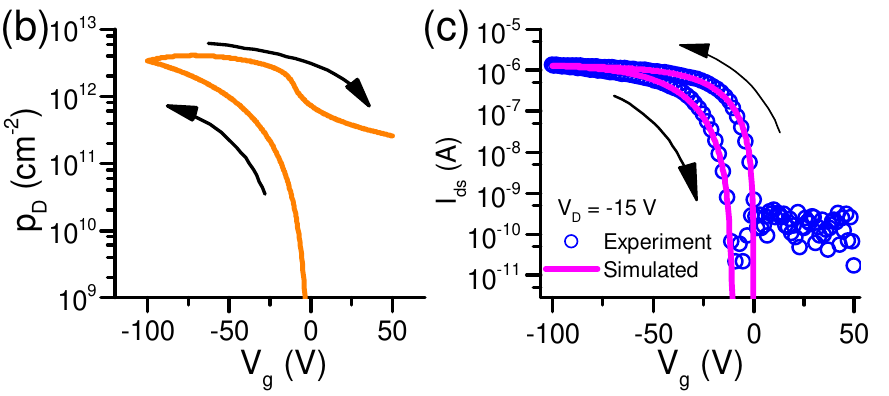}}
	\caption{(a) Schematic of the OTFT structure and layer composition. (b)The simulated DLT charge density versus gate bias. (c) The comparison of transfer characteristic between simulated and measured results. The arrows in (b) and (c) denote the sweeping direction of gate bias. } 
	\label{Fig.DeviceAndResults} 
\end{figure}

To examine our method, we fabricated a bottom-gate/ top-contact OTFT by using DPP-DTT as active layer. Fig. \ref{Fig.DeviceAndResults} (a) demonstrates the device structure. The architecture details are listed in Table. \ref{tab:table1}. The DPP-DTT is an excellent p-type organic semiconductor material with high carrier mobility and stable chemical structure \cite{RN449}. Its energy levels are listed in Table. \ref{tab:table1}. For the fabrication procedure, we first deposited its 5mg/ml chlorobenzene solution via spin-coating on a octyltrichlorosilane (OTS-8) treated Si/SiO$_2$ substrate. Then, 100nm gold layers were thermal evaporated using a shadow mask as source and drain electrodes. The electrical measurements for this OTFT was carried out by a Keithley 4200 semiconductor analyzer in atmosphere. The source contact was grounded and the drain bias was set to -15 V. The gate bias was first swept forwardly from 50 V to -100 V, then swept back to 50 V. The sweeping speed of gate bias was 10 V/s.

Both measured and simulated results are showed in Fig. \ref{Fig.DeviceAndResults} (b) and (c). 
All of simulation parameters are listed in Table \ref{tab:table1}. In Fig. \ref{Fig.DeviceAndResults}(b), the DLT charges increase while the device is driven to hole accumulation regime with enough negative gate bias. 
But as a nonequilibrium process, the trap concentration did not synchronize with the gate bias. The maximum point for DLT charges occurs around -60 V while sweeping the gate bias back to 50V. Fig. \ref{Fig.DeviceAndResults}(c) presented the simulated and measured transfer characteristic curves. Due to the accumulation of trapped DLT charges, the drain current in backward scan is smaller than that in forward scan. So, the Transfer curves present an anti-clockwise hysteresis loop and the threshold voltage shifts about -16 V in the backward scan. 

\section{Conclusion and Future Work}

We have combined theoretical framework of the trapping and detrapping pictures in organic semiconductors that comprise two Gaussian DOSs. 
Through the introducing of EES for DLT and EFS for HOMO, both trapping and detrapping expression were reduced to two simple product forms. 
This model is demonstrated in a FVM based device simulator. 
To verify this model, we fabricated a DPP-DTT based TFT with bottom-gate/ top-contact configuration.
The good agreement of the simulated and experimental results testify the reasonability of this model. 
A more detailed analysis on charge trapping is a subject for further investigation.

\section*{Acknowledgements}

This work is granted by the Natural Science Foundation of Guizhou Province (No. QKHJC-ZK[2021]YB329 and QKHJC-ZK[2021]YB018). We thanks Mr. Wenxuan Qiu for his language support.

\bibliography{bibliography}

\begin{thebibliography}{36}%
\makeatletter
\providecommand \@ifxundefined [1]{%
 \@ifx{#1\undefined}
}%
\providecommand \@ifnum [1]{%
 \ifnum #1\expandafter \@firstoftwo
 \else \expandafter \@secondoftwo
 \fi
}%
\providecommand \@ifx [1]{%
 \ifx #1\expandafter \@firstoftwo
 \else \expandafter \@secondoftwo
 \fi
}%
\providecommand \natexlab [1]{#1}%
\providecommand \enquote  [1]{``#1''}%
\providecommand \bibnamefont  [1]{#1}%
\providecommand \bibfnamefont [1]{#1}%
\providecommand \citenamefont [1]{#1}%
\providecommand \href@noop [0]{\@secondoftwo}%
\providecommand \href [0]{\begingroup \@sanitize@url \@href}%
\providecommand \@href[1]{\@@startlink{#1}\@@href}%
\providecommand \@@href[1]{\endgroup#1\@@endlink}%
\providecommand \@sanitize@url [0]{\catcode `\\12\catcode `\$12\catcode
  `\&12\catcode `\#12\catcode `\^12\catcode `\_12\catcode `\%12\relax}%
\providecommand \@@startlink[1]{}%
\providecommand \@@endlink[0]{}%
\providecommand \url  [0]{\begingroup\@sanitize@url \@url }%
\providecommand \@url [1]{\endgroup\@href {#1}{\urlprefix }}%
\providecommand \urlprefix  [0]{URL }%
\providecommand \Eprint [0]{\href }%
\providecommand \doibase [0]{https://doi.org/}%
\providecommand \selectlanguage [0]{\@gobble}%
\providecommand \bibinfo  [0]{\@secondoftwo}%
\providecommand \bibfield  [0]{\@secondoftwo}%
\providecommand \translation [1]{[#1]}%
\providecommand \BibitemOpen [0]{}%
\providecommand \bibitemStop [0]{}%
\providecommand \bibitemNoStop [0]{.\EOS\space}%
\providecommand \EOS [0]{\spacefactor3000\relax}%
\providecommand \BibitemShut  [1]{\csname bibitem#1\endcsname}%
\let\auto@bib@innerbib\@empty
\bibitem [{\citenamefont {Congreve}\ \emph {et~al.}(2013)\citenamefont
  {Congreve}, \citenamefont {Lee}, \citenamefont {Thompson}, \citenamefont
  {Hontz}, \citenamefont {Yost}, \citenamefont {Reusswig}, \citenamefont
  {Bahlke}, \citenamefont {Reineke}, \citenamefont {Van~Voorhis},\ and\
  \citenamefont {Baldo}}]{RN139}%
  \BibitemOpen
  \bibfield  {author} {\bibinfo {author} {\bibfnamefont {D.~N.}\ \bibnamefont
  {Congreve}}, \bibinfo {author} {\bibfnamefont {J.}~\bibnamefont {Lee}},
  \bibinfo {author} {\bibfnamefont {N.~J.}\ \bibnamefont {Thompson}}, \bibinfo
  {author} {\bibfnamefont {E.}~\bibnamefont {Hontz}}, \bibinfo {author}
  {\bibfnamefont {S.~R.}\ \bibnamefont {Yost}}, \bibinfo {author}
  {\bibfnamefont {P.~D.}\ \bibnamefont {Reusswig}}, \bibinfo {author}
  {\bibfnamefont {M.~E.}\ \bibnamefont {Bahlke}}, \bibinfo {author}
  {\bibfnamefont {S.}~\bibnamefont {Reineke}}, \bibinfo {author} {\bibfnamefont
  {T.}~\bibnamefont {Van~Voorhis}},\ and\ \bibinfo {author} {\bibfnamefont
  {M.~A.}\ \bibnamefont {Baldo}},\ }\bibfield  {title} {\enquote {\bibinfo
  {title} {External quantum efficiency above 100
  singlet-exciton-fission-based organic photovoltaic cell},}\ }\href
  {https://doi.org/10.1126/science.1232994} {\bibfield  {journal} {\bibinfo
  {journal} {Science}\ }\textbf {\bibinfo {volume} {340}},\ \bibinfo {pages}
  {334--337} (\bibinfo {year} {2013})}\BibitemShut {NoStop}%
\bibitem [{\citenamefont {Lin}\ \emph {et~al.}(2017)\citenamefont {Lin},
  \citenamefont {Guo}, \citenamefont {Chuang}, \citenamefont {Wang},
  \citenamefont {Wang}, \citenamefont {Liu}, \citenamefont {Hsu},\ and\
  \citenamefont {Jiang}}]{RN1500}%
  \BibitemOpen
  \bibfield  {author} {\bibinfo {author} {\bibfnamefont {F.~J.}\ \bibnamefont
  {Lin}}, \bibinfo {author} {\bibfnamefont {C.}~\bibnamefont {Guo}}, \bibinfo
  {author} {\bibfnamefont {W.~T.}\ \bibnamefont {Chuang}}, \bibinfo {author}
  {\bibfnamefont {C.~L.}\ \bibnamefont {Wang}}, \bibinfo {author}
  {\bibfnamefont {Q.}~\bibnamefont {Wang}}, \bibinfo {author} {\bibfnamefont
  {H.}~\bibnamefont {Liu}}, \bibinfo {author} {\bibfnamefont {C.~S.}\
  \bibnamefont {Hsu}},\ and\ \bibinfo {author} {\bibfnamefont {L.}~\bibnamefont
  {Jiang}},\ }\bibfield  {title} {\enquote {\bibinfo {title} {Directional
  solution coating by the chinese brush: A facile approach to improving
  molecular alignment for high-performance polymer tfts},}\ }\href
  {https://doi.org/10.1002/adma.201606987} {\bibfield  {journal} {\bibinfo
  {journal} {Adv Mater}\ }\textbf {\bibinfo {volume} {29}} (\bibinfo {year}
  {2017}),\ 10.1002/adma.201606987}\BibitemShut {NoStop}%
\bibitem [{\citenamefont {Zhao}\ \emph {et~al.}(2017)\citenamefont {Zhao},
  \citenamefont {Wang}, \citenamefont {Ni}, \citenamefont {Liu}, \citenamefont
  {Zhen}, \citenamefont {Zhang}, \citenamefont {Jiang}, \citenamefont {Li},
  \citenamefont {Dong},\ and\ \citenamefont {Hu}}]{RN546}%
  \BibitemOpen
  \bibfield  {author} {\bibinfo {author} {\bibfnamefont {Q.}~\bibnamefont
  {Zhao}}, \bibinfo {author} {\bibfnamefont {H.}~\bibnamefont {Wang}}, \bibinfo
  {author} {\bibfnamefont {Z.}~\bibnamefont {Ni}}, \bibinfo {author}
  {\bibfnamefont {J.}~\bibnamefont {Liu}}, \bibinfo {author} {\bibfnamefont
  {Y.}~\bibnamefont {Zhen}}, \bibinfo {author} {\bibfnamefont {X.}~\bibnamefont
  {Zhang}}, \bibinfo {author} {\bibfnamefont {L.}~\bibnamefont {Jiang}},
  \bibinfo {author} {\bibfnamefont {R.}~\bibnamefont {Li}}, \bibinfo {author}
  {\bibfnamefont {H.}~\bibnamefont {Dong}},\ and\ \bibinfo {author}
  {\bibfnamefont {W.}~\bibnamefont {Hu}},\ }\bibfield  {title} {\enquote
  {\bibinfo {title} {Organic ferroelectric-based 1t1t random access memory cell
  employing a common dielectric layer overcoming the half-selection problem},}\
  }\href {https://doi.org/10.1002/adma.201701907} {\bibfield  {journal}
  {\bibinfo  {journal} {Adv Mater}\ }\textbf {\bibinfo {volume} {29}} (\bibinfo
  {year} {2017}),\ 10.1002/adma.201701907}\BibitemShut {NoStop}%
\bibitem [{\citenamefont {Pandey}\ and\ \citenamefont {Nunzi}(2007)}]{Pandey}%
  \BibitemOpen
  \bibfield  {author} {\bibinfo {author} {\bibfnamefont {A.~K.}\ \bibnamefont
  {Pandey}}\ and\ \bibinfo {author} {\bibfnamefont {J.-M.}\ \bibnamefont
  {Nunzi}},\ }\bibfield  {title} {\enquote {\bibinfo {title} {Rubrene/fullerene
  heterostructures with a half-gap electroluminescence threshold and large
  photovoltage},}\ }\href {https://doi.org/10.1002/adma.200701052} {\bibfield
  {journal} {\bibinfo  {journal} {Advanced Materials}\ }\textbf {\bibinfo
  {volume} {19}},\ \bibinfo {pages} {3613--3617} (\bibinfo {year}
  {2007})}\BibitemShut {NoStop}%
\bibitem [{\citenamefont {Gross}\ \emph {et~al.}(2000)\citenamefont {Gross},
  \citenamefont {Müller}, \citenamefont {Nothofer}, \citenamefont {Scherf},
  \citenamefont {Neher}, \citenamefont {Bräuchle},\ and\ \citenamefont
  {Meerholz}}]{RN1938}%
  \BibitemOpen
  \bibfield  {author} {\bibinfo {author} {\bibfnamefont {M.}~\bibnamefont
  {Gross}}, \bibinfo {author} {\bibfnamefont {D.~C.}\ \bibnamefont {Müller}},
  \bibinfo {author} {\bibfnamefont {H.-G.}\ \bibnamefont {Nothofer}}, \bibinfo
  {author} {\bibfnamefont {U.}~\bibnamefont {Scherf}}, \bibinfo {author}
  {\bibfnamefont {D.}~\bibnamefont {Neher}}, \bibinfo {author} {\bibfnamefont
  {C.}~\bibnamefont {Bräuchle}},\ and\ \bibinfo {author} {\bibfnamefont
  {K.}~\bibnamefont {Meerholz}},\ }\bibfield  {title} {\enquote {\bibinfo
  {title} {Improving the performance of doped $\pi$-conjugated polymers for use
  in organic light-emitting diodes},}\ }\href
  {https://doi.org/10.1038/35015037} {\bibfield  {journal} {\bibinfo  {journal}
  {Nature}\ }\textbf {\bibinfo {volume} {405}},\ \bibinfo {pages} {661--665}
  (\bibinfo {year} {2000})}\BibitemShut {NoStop}%
\bibitem [{\citenamefont {Pfeiffer}\ \emph {et~al.}(2003)\citenamefont
  {Pfeiffer}, \citenamefont {Leo}, \citenamefont {Zhou}, \citenamefont {Huang},
  \citenamefont {Hofmann}, \citenamefont {Werner},\ and\ \citenamefont
  {Blochwitz-Nimoth}}]{RN1939}%
  \BibitemOpen
  \bibfield  {author} {\bibinfo {author} {\bibfnamefont {M.}~\bibnamefont
  {Pfeiffer}}, \bibinfo {author} {\bibfnamefont {K.}~\bibnamefont {Leo}},
  \bibinfo {author} {\bibfnamefont {X.}~\bibnamefont {Zhou}}, \bibinfo {author}
  {\bibfnamefont {J.}~\bibnamefont {Huang}}, \bibinfo {author} {\bibfnamefont
  {M.}~\bibnamefont {Hofmann}}, \bibinfo {author} {\bibfnamefont
  {A.}~\bibnamefont {Werner}},\ and\ \bibinfo {author} {\bibfnamefont
  {J.}~\bibnamefont {Blochwitz-Nimoth}},\ }\bibfield  {title} {\enquote
  {\bibinfo {title} {Doped organic semiconductors: Physics and application in
  light emitting diodes},}\ }\href
  {https://doi.org/https://doi.org/10.1016/j.orgel.2003.08.004} {\bibfield
  {journal} {\bibinfo  {journal} {Organic Electronics}\ }\textbf {\bibinfo
  {volume} {4}},\ \bibinfo {pages} {89--103} (\bibinfo {year} {2003})},\
  \bibinfo {note} {high Efficiency Light Emitters}\BibitemShut {NoStop}%
\bibitem [{\citenamefont {Hagfeldt}\ \emph {et~al.}(2010)\citenamefont
  {Hagfeldt}, \citenamefont {Boschloo}, \citenamefont {Sun}, \citenamefont
  {Kloo},\ and\ \citenamefont {Pettersson}}]{RN1937}%
  \BibitemOpen
  \bibfield  {author} {\bibinfo {author} {\bibfnamefont {A.}~\bibnamefont
  {Hagfeldt}}, \bibinfo {author} {\bibfnamefont {G.}~\bibnamefont {Boschloo}},
  \bibinfo {author} {\bibfnamefont {L.}~\bibnamefont {Sun}}, \bibinfo {author}
  {\bibfnamefont {L.}~\bibnamefont {Kloo}},\ and\ \bibinfo {author}
  {\bibfnamefont {H.}~\bibnamefont {Pettersson}},\ }\bibfield  {title}
  {\enquote {\bibinfo {title} {Dye-sensitized solar cells},}\ }\href
  {https://doi.org/10.1021/cr900356p} {\bibfield  {journal} {\bibinfo
  {journal} {Chemical reviews}\ }\textbf {\bibinfo {volume} {110}},\ \bibinfo
  {pages} {6595--6663} (\bibinfo {year} {2010})}\BibitemShut {NoStop}%
\bibitem [{\citenamefont {Paci}\ \emph {et~al.}(2006)\citenamefont {Paci},
  \citenamefont {Johnson}, \citenamefont {Chen}, \citenamefont {Rana},
  \citenamefont {Popovic}, \citenamefont {David}, \citenamefont {Nozik},
  \citenamefont {Ratner},\ and\ \citenamefont {Michl}}]{RN194}%
  \BibitemOpen
  \bibfield  {author} {\bibinfo {author} {\bibfnamefont {I.}~\bibnamefont
  {Paci}}, \bibinfo {author} {\bibfnamefont {J.~C.}\ \bibnamefont {Johnson}},
  \bibinfo {author} {\bibfnamefont {X.}~\bibnamefont {Chen}}, \bibinfo {author}
  {\bibfnamefont {G.}~\bibnamefont {Rana}}, \bibinfo {author} {\bibfnamefont
  {D.}~\bibnamefont {Popovic}}, \bibinfo {author} {\bibfnamefont {D.~E.}\
  \bibnamefont {David}}, \bibinfo {author} {\bibfnamefont {A.~J.}\ \bibnamefont
  {Nozik}}, \bibinfo {author} {\bibfnamefont {M.~A.}\ \bibnamefont {Ratner}},\
  and\ \bibinfo {author} {\bibfnamefont {J.}~\bibnamefont {Michl}},\ }\bibfield
   {title} {\enquote {\bibinfo {title} {Singlet fission for dye-sensitized
  solar cells: Can a suitable sensitizer be found?}}\ }\href
  {https://doi.org/10.1021/ja063980h} {\bibfield  {journal} {\bibinfo
  {journal} {Journal of the American Chemical Society}\ }\textbf {\bibinfo
  {volume} {128}},\ \bibinfo {pages} {16546--16553} (\bibinfo {year}
  {2006})}\BibitemShut {NoStop}%
\bibitem [{\citenamefont {Ya-Chin}, \citenamefont {Tsu-Jae},\ and\
  \citenamefont {Chenming}(2001)}]{RN1934}%
  \BibitemOpen
  \bibfield  {author} {\bibinfo {author} {\bibfnamefont {K.}~\bibnamefont
  {Ya-Chin}}, \bibinfo {author} {\bibfnamefont {K.}~\bibnamefont {Tsu-Jae}},\
  and\ \bibinfo {author} {\bibfnamefont {H.}~\bibnamefont {Chenming}},\
  }\bibfield  {title} {\enquote {\bibinfo {title} {Charge-trap memory device
  fabricated by oxidation of si/sub 1-x/ge/sub x},}\ }\href
  {https://doi.org/10.1109/16.915694} {\bibfield  {journal} {\bibinfo
  {journal} {IEEE Transactions on Electron Devices}\ }\textbf {\bibinfo
  {volume} {48}},\ \bibinfo {pages} {696--700} (\bibinfo {year}
  {2001})}\BibitemShut {NoStop}%
\bibitem [{\citenamefont {Lin}, \citenamefont {Lee},\ and\ \citenamefont
  {Chen}(2012)}]{RN1522}%
  \BibitemOpen
  \bibfield  {author} {\bibinfo {author} {\bibfnamefont {H.-W.}\ \bibnamefont
  {Lin}}, \bibinfo {author} {\bibfnamefont {W.-Y.}\ \bibnamefont {Lee}},\ and\
  \bibinfo {author} {\bibfnamefont {W.-C.}\ \bibnamefont {Chen}},\ }\bibfield
  {title} {\enquote {\bibinfo {title} {Selenophene-dpp donor–acceptor
  conjugated polymer for high performance ambipolar field effect transistor and
  nonvolatile memory applications},}\ }\href
  {https://doi.org/10.1039/c1jm14640h} {\bibfield  {journal} {\bibinfo
  {journal} {J. Mater. Chem.}\ }\textbf {\bibinfo {volume} {22}},\ \bibinfo
  {pages} {2120--2128} (\bibinfo {year} {2012})}\BibitemShut {NoStop}%
\bibitem [{\citenamefont {Haneef}, \citenamefont {Zeidell},\ and\ \citenamefont
  {Jurchescu}(2020)}]{RN1861}%
  \BibitemOpen
  \bibfield  {author} {\bibinfo {author} {\bibfnamefont {H.~F.}\ \bibnamefont
  {Haneef}}, \bibinfo {author} {\bibfnamefont {A.~M.}\ \bibnamefont
  {Zeidell}},\ and\ \bibinfo {author} {\bibfnamefont {O.~D.}\ \bibnamefont
  {Jurchescu}},\ }\bibfield  {title} {\enquote {\bibinfo {title} {Charge
  carrier traps in organic semiconductors: a review on the underlying physics
  and impact on electronic devices},}\ }\href
  {https://doi.org/10.1039/c9tc05695e} {\bibfield  {journal} {\bibinfo
  {journal} {Journal of Materials Chemistry C}\ }\textbf {\bibinfo {volume}
  {8}},\ \bibinfo {pages} {759--787} (\bibinfo {year} {2020})}\BibitemShut
  {NoStop}%
\bibitem [{\citenamefont {Street}\ \emph {et~al.}(2010)\citenamefont {Street},
  \citenamefont {Schoendorf}, \citenamefont {Roy},\ and\ \citenamefont
  {Lee}}]{RN301}%
  \BibitemOpen
  \bibfield  {author} {\bibinfo {author} {\bibfnamefont {R.~A.}\ \bibnamefont
  {Street}}, \bibinfo {author} {\bibfnamefont {M.}~\bibnamefont {Schoendorf}},
  \bibinfo {author} {\bibfnamefont {A.}~\bibnamefont {Roy}},\ and\ \bibinfo
  {author} {\bibfnamefont {J.~H.}\ \bibnamefont {Lee}},\ }\bibfield  {title}
  {\enquote {\bibinfo {title} {Interface state recombination in organic solar
  cells},}\ }\href {https://doi.org/10.1103/PhysRevB.81.205307} {\bibfield
  {journal} {\bibinfo  {journal} {Physical Review B}\ }\textbf {\bibinfo
  {volume} {81}},\ \bibinfo {pages} {205307} (\bibinfo {year}
  {2010})}\BibitemShut {NoStop}%
\bibitem [{\citenamefont {Zhan}\ \emph {et~al.}(2011)\citenamefont {Zhan},
  \citenamefont {Facchetti}, \citenamefont {Barlow}, \citenamefont {Marks},
  \citenamefont {Ratner}, \citenamefont {Wasielewski},\ and\ \citenamefont
  {Marder}}]{RN308}%
  \BibitemOpen
  \bibfield  {author} {\bibinfo {author} {\bibfnamefont {X.}~\bibnamefont
  {Zhan}}, \bibinfo {author} {\bibfnamefont {A.}~\bibnamefont {Facchetti}},
  \bibinfo {author} {\bibfnamefont {S.}~\bibnamefont {Barlow}}, \bibinfo
  {author} {\bibfnamefont {T.~J.}\ \bibnamefont {Marks}}, \bibinfo {author}
  {\bibfnamefont {M.~A.}\ \bibnamefont {Ratner}}, \bibinfo {author}
  {\bibfnamefont {M.~R.}\ \bibnamefont {Wasielewski}},\ and\ \bibinfo {author}
  {\bibfnamefont {S.~R.}\ \bibnamefont {Marder}},\ }\bibfield  {title}
  {\enquote {\bibinfo {title} {Rylene and related diimides for organic
  electronics},}\ }\href {https://doi.org/10.1002/adma.201001402} {\bibfield
  {journal} {\bibinfo  {journal} {Advanced Materials}\ }\textbf {\bibinfo
  {volume} {23}},\ \bibinfo {pages} {268--284} (\bibinfo {year}
  {2011})}\BibitemShut {NoStop}%
\bibitem [{\citenamefont {Bouhassoune}\ \emph {et~al.}(2009)\citenamefont
  {Bouhassoune}, \citenamefont {Mensfoort}, \citenamefont {Bobbert},\ and\
  \citenamefont {Coehoorn}}]{RN1906}%
  \BibitemOpen
  \bibfield  {author} {\bibinfo {author} {\bibfnamefont {M.}~\bibnamefont
  {Bouhassoune}}, \bibinfo {author} {\bibfnamefont {S.~L. M.~v.}\ \bibnamefont
  {Mensfoort}}, \bibinfo {author} {\bibfnamefont {P.~A.}\ \bibnamefont
  {Bobbert}},\ and\ \bibinfo {author} {\bibfnamefont {R.}~\bibnamefont
  {Coehoorn}},\ }\bibfield  {title} {\enquote {\bibinfo {title}
  {Carrier-density and field-dependent charge-carrier mobility in organic
  semiconductors with correlated gaussian disorder},}\ }\href
  {https://doi.org/10.1016/j.orgel.2009.01.005} {\bibfield  {journal} {\bibinfo
   {journal} {Organic Electronics}\ }\textbf {\bibinfo {volume} {10}},\
  \bibinfo {pages} {437--445} (\bibinfo {year} {2009})}\BibitemShut {NoStop}%
\bibitem [{\citenamefont {Coehoorn}\ \emph {et~al.}(2005)\citenamefont
  {Coehoorn}, \citenamefont {Pasveer}, \citenamefont {Bobbert},\ and\
  \citenamefont {Michels}}]{RN1907}%
  \BibitemOpen
  \bibfield  {author} {\bibinfo {author} {\bibfnamefont {R.}~\bibnamefont
  {Coehoorn}}, \bibinfo {author} {\bibfnamefont {W.~F.}\ \bibnamefont
  {Pasveer}}, \bibinfo {author} {\bibfnamefont {P.~A.}\ \bibnamefont
  {Bobbert}},\ and\ \bibinfo {author} {\bibfnamefont {M.~A.~J.}\ \bibnamefont
  {Michels}},\ }\bibfield  {title} {\enquote {\bibinfo {title} {Charge-carrier
  concentration dependence of the hopping mobility in organic materials with
  gaussian disorder},}\ }\href {https://doi.org/10.1103/PhysRevB.72.155206}
  {\bibfield  {journal} {\bibinfo  {journal} {Physical Review B}\ }\textbf
  {\bibinfo {volume} {72}} (\bibinfo {year} {2005}),\
  10.1103/PhysRevB.72.155206}\BibitemShut {NoStop}%
\bibitem [{\citenamefont {Chen}\ and\ \citenamefont {Xu}(2009)}]{RN1941}%
  \BibitemOpen
  \bibfield  {author} {\bibinfo {author} {\bibfnamefont {G.}~\bibnamefont
  {Chen}}\ and\ \bibinfo {author} {\bibfnamefont {Z.}~\bibnamefont {Xu}},\
  }\bibfield  {title} {\enquote {\bibinfo {title} {Charge trapping and
  detrapping in polymeric materials},}\ }\href
  {https://doi.org/10.1063/1.3273491} {\bibfield  {journal} {\bibinfo
  {journal} {Journal of Applied Physics}\ }\textbf {\bibinfo {volume} {106}}
  (\bibinfo {year} {2009}),\ 10.1063/1.3273491}\BibitemShut {NoStop}%
\bibitem [{\citenamefont {Alghamdi}, \citenamefont {Chen},\ and\ \citenamefont
  {Vaughan}()}]{RN1894}%
  \BibitemOpen
  \bibfield  {author} {\bibinfo {author} {\bibfnamefont {H.~A.}\ \bibnamefont
  {Alghamdi}}, \bibinfo {author} {\bibfnamefont {G.}~\bibnamefont {Chen}},\
  and\ \bibinfo {author} {\bibfnamefont {A.~S.}\ \bibnamefont {Vaughan}},\
  }\bibfield  {title} {\enquote {\bibinfo {title} {Simulation of the developed
  electro-thermal aging model based on trapping and detrapping process},}\
  }\href {https://doi.org/10.1109/CEIDP.2015.7352125} {\
  10.1109/CEIDP.2015.7352125}\BibitemShut {NoStop}%
\bibitem [{\citenamefont {Nicolai}, \citenamefont {Mandoc},\ and\ \citenamefont
  {Blom}(2011)}]{RN1926}%
  \BibitemOpen
  \bibfield  {author} {\bibinfo {author} {\bibfnamefont {H.~T.}\ \bibnamefont
  {Nicolai}}, \bibinfo {author} {\bibfnamefont {M.~M.}\ \bibnamefont
  {Mandoc}},\ and\ \bibinfo {author} {\bibfnamefont {P.~W.~M.}\ \bibnamefont
  {Blom}},\ }\bibfield  {title} {\enquote {\bibinfo {title} {Electron traps in
  semiconducting polymers: Exponential versus gaussian trap distribution},}\
  }\href {https://doi.org/10.1103/PhysRevB.83.195204} {\bibfield  {journal}
  {\bibinfo  {journal} {Physical Review B}\ }\textbf {\bibinfo {volume} {83}}
  (\bibinfo {year} {2011}),\ 10.1103/PhysRevB.83.195204}\BibitemShut {NoStop}%
\bibitem [{\citenamefont {Scheb}, \citenamefont {Zimmermann},\ and\
  \citenamefont {Jungemann}(2015)}]{RN1865}%
  \BibitemOpen
  \bibfield  {author} {\bibinfo {author} {\bibfnamefont {M.}~\bibnamefont
  {Scheb}}, \bibinfo {author} {\bibfnamefont {C.}~\bibnamefont {Zimmermann}},\
  and\ \bibinfo {author} {\bibfnamefont {C.}~\bibnamefont {Jungemann}},\
  }\bibfield  {title} {\enquote {\bibinfo {title} {Field-induced detrapping in
  doped organic semiconductors with gaussian disorder and different carrier
  localizations on host and guest sites},}\ }\href
  {https://doi.org/10.1103/PhysRevB.92.104201} {\bibfield  {journal} {\bibinfo
  {journal} {Physical Review B}\ }\textbf {\bibinfo {volume} {92}} (\bibinfo
  {year} {2015}),\ 10.1103/PhysRevB.92.104201}\BibitemShut {NoStop}%
\bibitem [{\citenamefont {Montero}\ and\ \citenamefont
  {Bisquert}(2011)}]{RN1920}%
  \BibitemOpen
  \bibfield  {author} {\bibinfo {author} {\bibfnamefont {J.~M.}\ \bibnamefont
  {Montero}}\ and\ \bibinfo {author} {\bibfnamefont {J.}~\bibnamefont
  {Bisquert}},\ }\bibfield  {title} {\enquote {\bibinfo {title} {Interpretation
  of trap-limited mobility in space-charge limited current in organic layers
  with exponential density of traps},}\ }\href
  {https://doi.org/10.1063/1.3622615} {\bibfield  {journal} {\bibinfo
  {journal} {Journal of Applied Physics}\ }\textbf {\bibinfo {volume} {110}}
  (\bibinfo {year} {2011}),\ 10.1063/1.3622615}\BibitemShut {NoStop}%
\bibitem [{\citenamefont {Germs}\ \emph {et~al.}(2011)\citenamefont {Germs},
  \citenamefont {van~der Holst}, \citenamefont {van Mensfoort}, \citenamefont
  {Bobbert},\ and\ \citenamefont {Coehoorn}}]{RN1946}%
  \BibitemOpen
  \bibfield  {author} {\bibinfo {author} {\bibfnamefont {W.~C.}\ \bibnamefont
  {Germs}}, \bibinfo {author} {\bibfnamefont {J.~J.~M.}\ \bibnamefont {van~der
  Holst}}, \bibinfo {author} {\bibfnamefont {S.~L.~M.}\ \bibnamefont {van
  Mensfoort}}, \bibinfo {author} {\bibfnamefont {P.~A.}\ \bibnamefont
  {Bobbert}},\ and\ \bibinfo {author} {\bibfnamefont {R.}~\bibnamefont
  {Coehoorn}},\ }\bibfield  {title} {\enquote {\bibinfo {title} {Modeling of
  the transient mobility in disordered organic semiconductors with a gaussian
  density of states},}\ }\href {https://doi.org/10.1103/PhysRevB.84.165210}
  {\bibfield  {journal} {\bibinfo  {journal} {Physical Review B}\ }\textbf
  {\bibinfo {volume} {84}} (\bibinfo {year} {2011}),\
  10.1103/PhysRevB.84.165210}\BibitemShut {NoStop}%
\bibitem [{\citenamefont {Miller}\ and\ \citenamefont
  {Abrahams}(1960)}]{RN1922}%
  \BibitemOpen
  \bibfield  {author} {\bibinfo {author} {\bibfnamefont {A.}~\bibnamefont
  {Miller}}\ and\ \bibinfo {author} {\bibfnamefont {E.}~\bibnamefont
  {Abrahams}},\ }\bibfield  {title} {\enquote {\bibinfo {title} {Impurity
  conduction at low concentrations},}\ }\href
  {https://doi.org/10.1103/PhysRev.120.745} {\bibfield  {journal} {\bibinfo
  {journal} {Physical Review}\ }\textbf {\bibinfo {volume} {120}},\ \bibinfo
  {pages} {745--755} (\bibinfo {year} {1960})}\BibitemShut {NoStop}%
\bibitem [{\citenamefont {Lei}\ \emph {et~al.}(2016)\citenamefont {Lei},
  \citenamefont {Deng}, \citenamefont {Lin}, \citenamefont {Zheng},
  \citenamefont {Zhu},\ and\ \citenamefont {Ong}}]{RN449}%
  \BibitemOpen
  \bibfield  {author} {\bibinfo {author} {\bibfnamefont {Y.}~\bibnamefont
  {Lei}}, \bibinfo {author} {\bibfnamefont {P.}~\bibnamefont {Deng}}, \bibinfo
  {author} {\bibfnamefont {M.}~\bibnamefont {Lin}}, \bibinfo {author}
  {\bibfnamefont {X.}~\bibnamefont {Zheng}}, \bibinfo {author} {\bibfnamefont
  {F.}~\bibnamefont {Zhu}},\ and\ \bibinfo {author} {\bibfnamefont {B.~S.}\
  \bibnamefont {Ong}},\ }\bibfield  {title} {\enquote {\bibinfo {title}
  {Enhancing crystalline structural orders of polymer semiconductors for
  efficient charge transport via polymer-matrix-mediated molecular
  self-assembly},}\ }\href {https://doi.org/10.1002/adma.201600580} {\bibfield
  {journal} {\bibinfo  {journal} {Adv Mater}\ }\textbf {\bibinfo {volume}
  {28}},\ \bibinfo {pages} {6687--94} (\bibinfo {year} {2016})}\BibitemShut
  {NoStop}%
\bibitem [{\citenamefont {{DEVSIM LLC}}()}]{DEVSIM}%
  \BibitemOpen
  \bibfield  {author} {\bibinfo {author} {\bibnamefont {{DEVSIM LLC}}},\
  }\href@noop {} {\enquote {\bibinfo {title} {{DEVSIM} {TCAD} semiconductor
  device simulator},}\ }\bibinfo {howpublished}
  {\url{https://devsim.org}}\BibitemShut {NoStop}%
\bibitem [{\citenamefont {Kimura}\ \emph {et~al.}(2008)\citenamefont {Kimura},
  \citenamefont {Nakanishi}, \citenamefont {Nomura}, \citenamefont {Kamiya},\
  and\ \citenamefont {Hosono}}]{RN1796}%
  \BibitemOpen
  \bibfield  {author} {\bibinfo {author} {\bibfnamefont {M.}~\bibnamefont
  {Kimura}}, \bibinfo {author} {\bibfnamefont {T.}~\bibnamefont {Nakanishi}},
  \bibinfo {author} {\bibfnamefont {K.}~\bibnamefont {Nomura}}, \bibinfo
  {author} {\bibfnamefont {T.}~\bibnamefont {Kamiya}},\ and\ \bibinfo {author}
  {\bibfnamefont {H.}~\bibnamefont {Hosono}},\ }\bibfield  {title} {\enquote
  {\bibinfo {title} {Trap densities in amorphous-ingazno4 thin-film
  transistors},}\ }\href {https://doi.org/10.1063/1.2904704} {\bibfield
  {journal} {\bibinfo  {journal} {Applied Physics Letters}\ }\textbf {\bibinfo
  {volume} {92}} (\bibinfo {year} {2008}),\ 10.1063/1.2904704}\BibitemShut
  {NoStop}%
\bibitem [{\citenamefont {Xu}\ \emph {et~al.}(2016)\citenamefont {Xu},
  \citenamefont {Zhai}, \citenamefont {Tang}, \citenamefont {Qiu},
  \citenamefont {Liu}, \citenamefont {Rong}, \citenamefont {Pang},
  \citenamefont {Jiang}, \citenamefont {Xiao}, \citenamefont {Zhong},
  \citenamefont {Mi}, \citenamefont {Fan},\ and\ \citenamefont
  {Huang}}]{RN1526}%
  \BibitemOpen
  \bibfield  {author} {\bibinfo {author} {\bibfnamefont {H.}~\bibnamefont
  {Xu}}, \bibinfo {author} {\bibfnamefont {W.-J.}\ \bibnamefont {Zhai}},
  \bibinfo {author} {\bibfnamefont {C.}~\bibnamefont {Tang}}, \bibinfo {author}
  {\bibfnamefont {S.-Y.}\ \bibnamefont {Qiu}}, \bibinfo {author} {\bibfnamefont
  {R.-L.}\ \bibnamefont {Liu}}, \bibinfo {author} {\bibfnamefont
  {Z.}~\bibnamefont {Rong}}, \bibinfo {author} {\bibfnamefont {Z.-Q.}\
  \bibnamefont {Pang}}, \bibinfo {author} {\bibfnamefont {B.}~\bibnamefont
  {Jiang}}, \bibinfo {author} {\bibfnamefont {J.}~\bibnamefont {Xiao}},
  \bibinfo {author} {\bibfnamefont {C.}~\bibnamefont {Zhong}}, \bibinfo
  {author} {\bibfnamefont {B.-X.}\ \bibnamefont {Mi}}, \bibinfo {author}
  {\bibfnamefont {Q.-L.}\ \bibnamefont {Fan}},\ and\ \bibinfo {author}
  {\bibfnamefont {W.}~\bibnamefont {Huang}},\ }\bibfield  {title} {\enquote
  {\bibinfo {title} {Thickness dependence of carrier mobility and the interface
  trap free energy investigated by impedance spectroscopy in organic
  semiconductors},}\ }\href {https://doi.org/10.1021/acs.jpcc.6b03964}
  {\bibfield  {journal} {\bibinfo  {journal} {The Journal of Physical Chemistry
  C}\ }\textbf {\bibinfo {volume} {120}},\ \bibinfo {pages} {17184--17189}
  (\bibinfo {year} {2016})}\BibitemShut {NoStop}%
\bibitem [{\citenamefont {Paasch}\ and\ \citenamefont
  {Scheinert}(2010)}]{Paasch}%
  \BibitemOpen
  \bibfield  {author} {\bibinfo {author} {\bibfnamefont {G.}~\bibnamefont
  {Paasch}}\ and\ \bibinfo {author} {\bibfnamefont {S.}~\bibnamefont
  {Scheinert}},\ }\bibfield  {title} {\enquote {\bibinfo {title} {Charge
  carrier density of organics with gaussian density of states: Analytical
  approximation for the gauss–fermi integral},}\ }\href
  {https://doi.org/10.1063/1.3374475} {\bibfield  {journal} {\bibinfo
  {journal} {Journal of Applied Physics}\ }\textbf {\bibinfo {volume} {107}}
  (\bibinfo {year} {2010}),\ 10.1063/1.3374475}\BibitemShut {NoStop}%
\bibitem [{\citenamefont {Neamen}(2003)}]{RN1925}%
  \BibitemOpen
  \bibfield  {author} {\bibinfo {author} {\bibfnamefont {D.~A.}\ \bibnamefont
  {Neamen}},\ }\href@noop {} {\emph {\bibinfo {title} {Semiconductor physics
  and devices}}}\ (\bibinfo  {publisher} {McGraw-Hill higher education},\
  \bibinfo {year} {2003})\BibitemShut {NoStop}%
\bibitem [{\citenamefont {Oelerich}, \citenamefont {Huemmer},\ and\
  \citenamefont {Baranovskii}(2012)}]{RN1877}%
  \BibitemOpen
  \bibfield  {author} {\bibinfo {author} {\bibfnamefont {J.~O.}\ \bibnamefont
  {Oelerich}}, \bibinfo {author} {\bibfnamefont {D.}~\bibnamefont {Huemmer}},\
  and\ \bibinfo {author} {\bibfnamefont {S.~D.}\ \bibnamefont {Baranovskii}},\
  }\bibfield  {title} {\enquote {\bibinfo {title} {How to find out the density
  of states in disordered organic semiconductors},}\ }\href
  {https://doi.org/10.1103/PhysRevLett.108.226403} {\bibfield  {journal}
  {\bibinfo  {journal} {Phys Rev Lett}\ }\textbf {\bibinfo {volume} {108}},\
  \bibinfo {pages} {226403} (\bibinfo {year} {2012})}\BibitemShut {NoStop}%
\bibitem [{\citenamefont {Knapp}\ \emph {et~al.}(2010)\citenamefont {Knapp},
  \citenamefont {Häusermann}, \citenamefont {Schwarzenbach},\ and\
  \citenamefont {Ruhstaller}}]{RN1898}%
  \BibitemOpen
  \bibfield  {author} {\bibinfo {author} {\bibfnamefont {E.}~\bibnamefont
  {Knapp}}, \bibinfo {author} {\bibfnamefont {R.}~\bibnamefont {Häusermann}},
  \bibinfo {author} {\bibfnamefont {H.~U.}\ \bibnamefont {Schwarzenbach}},\
  and\ \bibinfo {author} {\bibfnamefont {B.}~\bibnamefont {Ruhstaller}},\
  }\bibfield  {title} {\enquote {\bibinfo {title} {Numerical simulation of
  charge transport in disordered organic semiconductor devices},}\ }\href
  {https://doi.org/10.1063/1.3475505} {\bibfield  {journal} {\bibinfo
  {journal} {Journal of Applied Physics}\ }\textbf {\bibinfo {volume} {108}}
  (\bibinfo {year} {2010}),\ 10.1063/1.3475505}\BibitemShut {NoStop}%
\bibitem [{\citenamefont {Boubaker}\ \emph {et~al.}(2017)\citenamefont
  {Boubaker}, \citenamefont {Hafsi}, \citenamefont {Lmimouni},\ and\
  \citenamefont {Kalboussi}}]{RN1773}%
  \BibitemOpen
  \bibfield  {author} {\bibinfo {author} {\bibfnamefont {A.}~\bibnamefont
  {Boubaker}}, \bibinfo {author} {\bibfnamefont {B.}~\bibnamefont {Hafsi}},
  \bibinfo {author} {\bibfnamefont {K.}~\bibnamefont {Lmimouni}},\ and\
  \bibinfo {author} {\bibfnamefont {A.}~\bibnamefont {Kalboussi}},\ }\bibfield
  {title} {\enquote {\bibinfo {title} {A comparative tcad simulations of a
  p-and n-type organic field effect transistors: field-dependent mobility, bulk
  and interface traps models},}\ }\href
  {https://doi.org/10.1007/s10854-017-6480-y} {\bibfield  {journal} {\bibinfo
  {journal} {Journal of Materials Science: Materials in Electronics}\ }\textbf
  {\bibinfo {volume} {28}},\ \bibinfo {pages} {7834--7843} (\bibinfo {year}
  {2017})}\BibitemShut {NoStop}%
\bibitem [{\citenamefont {Baranovskii}(2018)}]{RN1873}%
  \BibitemOpen
  \bibfield  {author} {\bibinfo {author} {\bibfnamefont {S.~D.}\ \bibnamefont
  {Baranovskii}},\ }\bibfield  {title} {\enquote {\bibinfo {title} {Mott
  lecture: Description of charge transport in disordered organic
  semiconductors: Analytical theories and computer simulations},}\ }\href
  {https://doi.org/10.1002/pssa.201700676} {\bibfield  {journal} {\bibinfo
  {journal} {physica status solidi (a)}\ }\textbf {\bibinfo {volume} {215}}
  (\bibinfo {year} {2018}),\ 10.1002/pssa.201700676}\BibitemShut {NoStop}%
\bibitem [{\citenamefont {Li}\ \emph {et~al.}(2012)\citenamefont {Li},
  \citenamefont {Zhao}, \citenamefont {Tan}, \citenamefont {Guo}, \citenamefont
  {Di}, \citenamefont {Yu}, \citenamefont {Liu}, \citenamefont {Lin},
  \citenamefont {Lim}, \citenamefont {Zhou}, \citenamefont {Su},\ and\
  \citenamefont {Ong}}]{RN1550}%
  \BibitemOpen
  \bibfield  {author} {\bibinfo {author} {\bibfnamefont {J.}~\bibnamefont
  {Li}}, \bibinfo {author} {\bibfnamefont {Y.}~\bibnamefont {Zhao}}, \bibinfo
  {author} {\bibfnamefont {H.~S.}\ \bibnamefont {Tan}}, \bibinfo {author}
  {\bibfnamefont {Y.}~\bibnamefont {Guo}}, \bibinfo {author} {\bibfnamefont
  {C.~A.}\ \bibnamefont {Di}}, \bibinfo {author} {\bibfnamefont
  {G.}~\bibnamefont {Yu}}, \bibinfo {author} {\bibfnamefont {Y.}~\bibnamefont
  {Liu}}, \bibinfo {author} {\bibfnamefont {M.}~\bibnamefont {Lin}}, \bibinfo
  {author} {\bibfnamefont {S.~H.}\ \bibnamefont {Lim}}, \bibinfo {author}
  {\bibfnamefont {Y.}~\bibnamefont {Zhou}}, \bibinfo {author} {\bibfnamefont
  {H.}~\bibnamefont {Su}},\ and\ \bibinfo {author} {\bibfnamefont {B.~S.}\
  \bibnamefont {Ong}},\ }\bibfield  {title} {\enquote {\bibinfo {title} {A
  stable solution-processed polymer semiconductor with record high-mobility for
  printed transistors},}\ }\href {https://doi.org/10.1038/srep00754} {\bibfield
   {journal} {\bibinfo  {journal} {Sci Rep}\ }\textbf {\bibinfo {volume} {2}},\
  \bibinfo {pages} {754} (\bibinfo {year} {2012})}\BibitemShut {NoStop}%
\bibitem [{\citenamefont {Ishibashi}(1994)}]{RN1932}%
  \BibitemOpen
  \bibfield  {author} {\bibinfo {author} {\bibfnamefont {T.}~\bibnamefont
  {Ishibashi}},\ }\enquote {\bibinfo {title} {Chapter 5 - gaas-based and
  inp-based heterostructure bipolar transistors},}\ in\ \href
  {https://doi.org/https://doi.org/10.1016/S0080-8784(08)62479-5} {\emph
  {\bibinfo {booktitle} {Semiconductors and Semimetals}}},\ Vol.~\bibinfo
  {volume} {41},\ \bibinfo {editor} {edited by\ \bibinfo {editor}
  {\bibfnamefont {R.~A.}\ \bibnamefont {Kiehl}}\ and\ \bibinfo {editor}
  {\bibfnamefont {T.~C. L.~G.}\ \bibnamefont {Sollner}}}\ (\bibinfo
  {publisher} {Elsevier},\ \bibinfo {year} {1994})\ pp.\ \bibinfo {pages}
  {291--358}\BibitemShut {NoStop}%
\bibitem [{\citenamefont {Ghittorelli}\ \emph {et~al.}(2017)\citenamefont
  {Ghittorelli}, \citenamefont {Lenz}, \citenamefont {Sharifi~Dehsari},
  \citenamefont {Zhao}, \citenamefont {Asadi}, \citenamefont {Blom},
  \citenamefont {Kovacs-Vajna}, \citenamefont {de~Leeuw},\ and\ \citenamefont
  {Torricelli}}]{RN537}%
  \BibitemOpen
  \bibfield  {author} {\bibinfo {author} {\bibfnamefont {M.}~\bibnamefont
  {Ghittorelli}}, \bibinfo {author} {\bibfnamefont {T.}~\bibnamefont {Lenz}},
  \bibinfo {author} {\bibfnamefont {H.}~\bibnamefont {Sharifi~Dehsari}},
  \bibinfo {author} {\bibfnamefont {D.}~\bibnamefont {Zhao}}, \bibinfo {author}
  {\bibfnamefont {K.}~\bibnamefont {Asadi}}, \bibinfo {author} {\bibfnamefont
  {P.~W.~M.}\ \bibnamefont {Blom}}, \bibinfo {author} {\bibfnamefont {Z.~M.}\
  \bibnamefont {Kovacs-Vajna}}, \bibinfo {author} {\bibfnamefont {D.~M.}\
  \bibnamefont {de~Leeuw}},\ and\ \bibinfo {author} {\bibfnamefont
  {F.}~\bibnamefont {Torricelli}},\ }\bibfield  {title} {\enquote {\bibinfo
  {title} {Quantum tunnelling and charge accumulation in organic ferroelectric
  memory diodes},}\ }\href {https://doi.org/10.1038/ncomms15841} {\bibfield
  {journal} {\bibinfo  {journal} {Nature Communications}\ }\textbf {\bibinfo
  {volume} {8}},\ \bibinfo {pages} {15741} (\bibinfo {year}
  {2017})}\BibitemShut {NoStop}%
\bibitem [{RN1(2006)}]{RN1933}%
  \BibitemOpen
  \enquote {\bibinfo {title} {Metal-insulator-semiconductor capacitors},}\ in\
  \href {https://doi.org/https://doi.org/10.1002/9780470068328.ch4} {\emph
  {\bibinfo {booktitle} {Physics of Semiconductor Devices}}}\ (\bibinfo {year}
  {2006})\ pp.\ \bibinfo {pages} {197--240}\BibitemShut {NoStop}%
\end{thebibliography}%

\end{document}